\documentclass[%
 reprint, aip,
amsmath,amssymb,
floatfix
]{revtex4-2}

\usepackage[subpreambles=true]{standalone}
\usepackage{import, inputenc} % import standalones
\usepackage{amsmath, amssymb,amsfonts, caption, graphicx, epstopdf, amsthm, hyperref, subcaption}

\usepackage{braket,tensor, tikz,multirow, bm}
\usepackage{placeins, pgfplots, pgfplotstable}
\usepackage{array}	% for \newcolumntype macro
\usetikzlibrary{intersections, arrows,decorations.pathmorphing, snakes, shapes, positioning, patterns, decorations.text,decorations.pathreplacing}

\usepackage{breakcites}

\usepackage{etoolbox}
\makeatletter
\def\@email#1#2{%
 \endgroup
 \patchcmd{\titleblock@produce}
  {\frontmatter@RRAPformat}
  {\frontmatter@RRAPformat{\produce@RRAP{*#1\href{mailto:#2}{#2}}}\frontmatter@RRAPformat}
  {}{}
}%
\makeatother

\begin{document}
\title[An improved synthetic signal injection routine for the Haloscope At Yale Sensitive To Axion Cold dark matter (HAYSTAC)]{An improved synthetic signal injection routine for the Haloscope At Yale Sensitive To Axion Cold dark matter (HAYSTAC)}

\author{Yuqi Zhu}
\thanks{Present address: Stanford University, Stanford, CA 94305, USA}
\affiliation{Department of Physics, Yale University, New Haven, Connecticut 06520, USA}
\affiliation{Wright Laboratory, Department of Physics, Yale University, New Haven, Connecticut 06520, USA}

\author{M. J.\ Jewell}
% \thanks{These authors contributed equally to this work.}
\affiliation{Department of Physics, Yale University, New Haven, Connecticut 06520, USA}
\affiliation{Wright Laboratory, Department of Physics, Yale University, New Haven, Connecticut 06520, USA}

\author{Claire Laffan}
\affiliation{Department of Physics, Yale University, New Haven, Connecticut 06520, USA}
\affiliation{Wright Laboratory, Department of Physics, Yale University, New Haven, Connecticut 06520, USA}

\author{Xiran Bai}
\affiliation{Department of Physics, Yale University, New Haven, Connecticut 06520, USA}
\affiliation{Wright Laboratory, Department of Physics, Yale University, New Haven, Connecticut 06520, USA}

\author{Sumita Ghosh}
\affiliation{Wright Laboratory, Department of Physics, Yale University, New Haven, Connecticut 06520, USA}
\affiliation{Department of Applied Physics, Yale University, New Haven, Connecticut 06520, USA}

\author{Eleanor Graham}
\affiliation{Department of Physics, Yale University, New Haven, Connecticut 06520, USA}
\affiliation{Wright Laboratory, Department of Physics, Yale University, New Haven, Connecticut 06520, USA}

\author{S. B.\ Cahn}
\affiliation{Department of Physics, Yale University, New Haven, Connecticut 06520, USA}
\affiliation{Wright Laboratory, Department of Physics, Yale University, New Haven, Connecticut 06520, USA}

\author{Reina H.\ Maruyama}\thanks{The author to whom correspondence may be addressed: reina.maruyama@yale.edu}
% \email{reina.maruyama@yale.edu}
\affiliation{Department of Physics, Yale University, New Haven, Connecticut 06520, USA}
\affiliation{Wright Laboratory, Department of Physics, Yale University, New Haven, Connecticut 06520, USA}

\author{S. K.\ Lamoreaux}
\affiliation{Department of Physics, Yale University, New Haven, Connecticut 06520, USA}
\affiliation{Wright Laboratory, Department of Physics, Yale University, New Haven, Connecticut 06520, USA}

\date{\today}
\begin{abstract}
Microwave cavity haloscopes are among the most sensitive direct detection experiments searching for dark matter axions via their coupling to photons. When the power of the expected microwave signal due to axion-photon conversion is on the order of $10^{-24}$~W, having the ability to validate the detector response and analysis procedure by injecting realistic synthetic axion signals becomes helpful. Here we present a method based on frequency hopping spread spectrum for synthesizing axion signals in a microwave cavity haloscope experiment. It allows us to generate a narrow and asymmetric shape in frequency space that mimics an axion's spectral distribution, which is derived from a Maxwell-Boltzmann distribution. In addition, we show that the synthetic axion's power can be calibrated with reference to the system noise. Compared to the synthetic axion injection in HAYSTAC phase I, we demonstrated synthetic signal injection with a more realistic lineshape and calibrated power.
\end{abstract}
\maketitle

\section{Introduction}
The axion is a well-motivated solution to the strong charge–parity problem in quantum chromodynamics and is also a dark matter candidate~\cite{Peccei, Peccei2, Weinberg1978, Wilczek1978}. The most sensitive direct detection experiments by far are based on the microwave cavity haloscope technique~\cite{Sikivie1983}. In a haloscope experiment, the axions would convert into photons of equal energy inside a resonant microwave cavity permeated by a strong magnetic field. The energy of each axion-converted photon would be equal to the axion's total energy---the sum of its rest-mass energy $m_a$ and kinetic energy. Therefore, the lineshape of the resulting microwave signal appearing in a haloscope experiment would inherit the axion's kinetic energy distribution. Suppose $\nu_a \!= \! m_a/(2\pi)$ is the axion mass in SI-frequency units, we denote the spectral distribution of an axion with mass $m_a$ as $f_{\nu_a}(\nu)$. The distribution $f_{\nu_a}(\nu)$ encodes axion's properties derived from the pseudo-isothermal halo model~\cite{Jimenez2003}, and accounts for the modulation due to the Earth's rotation around the center of the galaxy~\cite{Turner1990, Brubaker2017procedure}.  
Here we describe a method to synthesize a microwave signal whose lineshape resembles that from an expected axion signal. 
Injecting and detecting such synthesized signals into the detector allows us to characterize the detector’s response, validate the analysis procedure~\cite{HAYSTAC2017, Brubaker2017procedure, ALKENANY2017}, and perform a blind analysis~\cite{admx2021}.

Our method for synthesizing axion signals is inspired by a patent for radio frequency (rf) hopping by Markey and Antheil~\cite{Lamarrpatent}; it is related to the frequency hopping spread spectrum (FHSS) technique, which has been applied in military and wireless communication to prevent interception and reduce interference. In our experiment, this technique allows us to generate an axion's spectrum by hopping between a large number of random rf tones sampled from the axion's spectral distribution. This way we can produce the spectral spread associated with the axion's kinetic energy distribution. As will become more apparent later, the axion's lineshape function is asymmetric and narrow. The frequency hopping method allows us to overcome the technical difficulties associated with signal shaping in the frequency domain. In principle, this technique can be used to synthesize any generic spectral shapes. 
It is also a realistic simulation: each axion's energy is a random variable that follows $f_{\nu_a}(\nu)$ and the observed lineshape is an ensemble-averaged result.

In this note, we describe the methods involved in signal injection in Sec.~\ref{sec: method} and provide a summary and outlook in Sec.~\ref{sec: outlook}. 
 
\section{Methods \label{sec: method}}
\begin{figure*}
    \centering
    \resizebox{.8\textwidth}{!}{
    % \import{}{tikzfigures/schematic2}
    \begin{tikzpicture}[
  scale=1.8, dev/.style ={rectangle, draw=black!50, ultra thick, text width=4em, align=center, rounded corners, minimum height=2em, text=black!50, font=\bf},   l/.style={ultra thick,>=stealth, blue1,->}, dev2/.style ={rectangle, fill=black!30, ultra thick, text width=4em, align=center, rounded corners, minimum height=2em, text=white, font=\bf},  pc/.style={ text width=4 em, align=center, minimum height=2em}]
  
\definecolor{blue1}{HTML}{1a6eff}
\pgfmathsetmacro\h{1.5} %horizontal sep.  
\pgfmathsetmacro\v{-1.5} %vertical sep.  
\path  (-\h, 0) node[dev2] (pc2) {PC2}
(0, 0) node[dev2](psg2){PSG2}
(\h*1,0) node[dev2] (rfsw) {rf switch}
(\h*2,0) node[dev2] (splitter) {power splitter}
(\h*3,-\v/2) node[dev, text width=3em] (usbdac) {USB DAC}
(2*\h,\v/2) node[dev](cavity){Cavity}
(2*\h,1.5*\v) node[dev](jpa){AMPs}
(.5*\h,\v) node[dev,text width=2.5em](vts){VTS}
(2*\h,\v)node[dev, circle, text width=1em](circ){}
(\h,2*\v) node[dev] (psg1) {PSG1}
% (\h,2*\v) node[dev]  (div) {freq. divider}
(2*\h,2*\v) node[black!50, inner sep=0](mixer){\Huge $\bm \otimes$}
(2.7*\h,2*\v) node[dev, text width=3em](lpf){LPF}
(3.7*\h,2*\v) node[dev](dac){ADC}
(1.5*\h,\v) node[dev, text width=2.5em] (sq) {SQZ};
% circulator arrow 
\draw[ultra thick,black!50] (circ.west)+(.2em,0) arc (180:-90:.4em) node[left] (x){};
\draw[->, >=stealth,ultra thick,black!50] (x)--++(0.01,0);
\node[right, gray] at (circ.east) {\bf circulator};
%other arrows
\draw[l, black, dashed,<->] (pc2.east) -- (psg2.west) node[midway, above]{VISA};
\draw[l] (psg2.east) -- (rfsw.west);
\draw[l,  black, dashed] (usbdac.west)-|(rfsw.north) node[midway, above]{TTL};
\draw[l,->]  (rfsw.east)--(splitter.west);
\draw[l,<-]  (splitter.east)--++(\h/2,0) node[right]{other tx in};
\draw[l, black, <->, dashed] (dac.north) --++(0,-.3*\v) node[dev,solid, above](pc1) {PC1};
\draw[l]  (splitter.south)--(cavity.north);
\draw[l] (cavity.south)--(circ.north);
\draw[l] (circ.south)--(jpa.north);
\draw[l] (jpa.south)--(mixer.north)node[right,midway]{RF};
% dotted
\draw[l, loosely dotted] (vts.east)--(sq.west) node[midway, above]{calibration};
\draw[l, loosely dotted] (sq.9)--(circ.171);
\draw[l] (sq.east)--(circ.west);
\draw[l, loosely dotted] (circ.110)--(cavity.250);

\draw[l] (psg1.east) -- (mixer.west)node[above,midway]{LO};
% \draw[l] (div.east) -- (mixer.west) node[above,midway]{LO};
\draw[l] (mixer.east) --(lpf.west);
\draw[l] (lpf.east) --(dac.west);
\draw[l, black, dashed, <->] (pc1.north)|-(usbdac.east);
\node[above, text=black!50] at (cavity.20) {weak port};
\node[right, yshift=-3pt, text=black!50] at (cavity.280) {receiver's port};
\end{tikzpicture}
    }
    \caption{Schematic. New components added for hardware injection are shown as gray nodes; the rest already exists as part of the experiment. Blue arrows and black dashed arrows indicate analog and digital signals respectively. PC2 controls PSG2 to inject a number of rf tones whose frequencies are sampled from $f_{\nu_a}(\nu)$ to simulate axions with mass $\nu_a$. The synthetic axion signals are injected through the weak port of the cavity and then measured and amplified by the receiver chain consisting of two amplifiers (AMPs): a JPA followed by a High-electron-mobility transistor. 
    Subsequently, the amplified signals are down-converted, mixed with the local oscillator (LO) signal provided by PSG1,  and then filtered by the low-pass filter (LPF). The time-series signal collected by the Analog/Digital Converter (ADC) is then Fourier transformed into an intermediate-frequency (IF) spectrum. The VTS helps us characterize the system's total noise by comparing the cavity's spectrum to blackbody spectra at controlled temperatures. The signal from the VTS is only coupled in during a calibration measurement, indicated by the dotted blue arrows.
    }
    \label{fig: schematic}
\end{figure*}
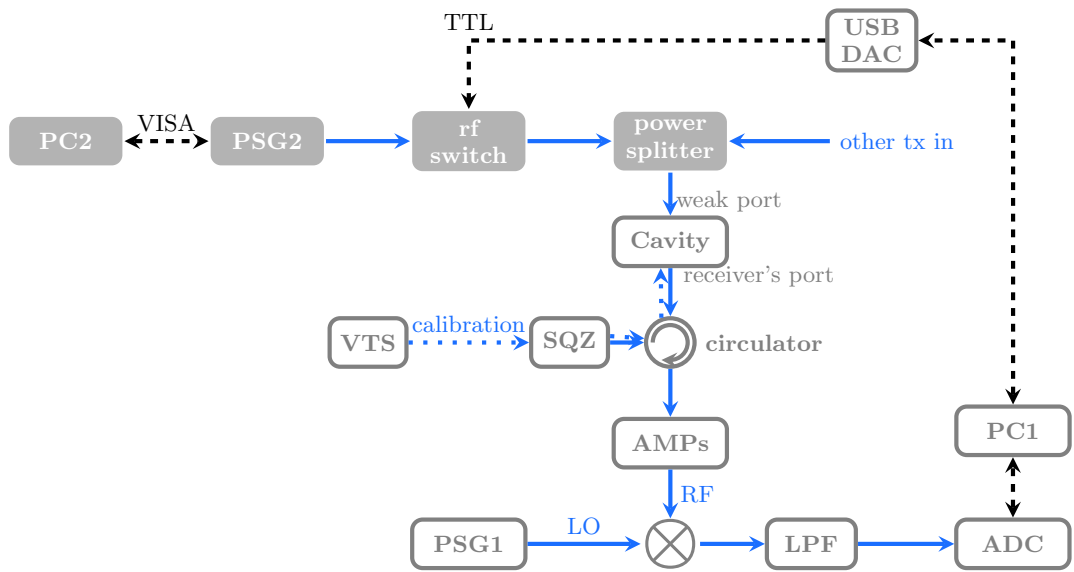

The synthetic axion signal is injected as a transmission (tx) input through the weak port of the cavity and subsequently coupled out from the receiver's port. Fig.~\ref{fig: schematic} shows the schematic of synthetic signal injection. 
The main device that synthesizes axion signals is an analog signal generator (model:~Agilent E8257D), labeled as PSG2. 
In between PSG2 and the cavity's weak port, an rf switch (model:~ZFSWA2-63DR+) and a power splitter/combiner (model:~ZX10-2-71-S+) are added in this order. 
The rf switch allows us to interrupt signal injection from PC1 via a USB DAC (model:~USB-6009-NI). For instance, the injection signals are blocked during calibration measurements between axion datasets, as they would bias the noise measurement. % and its input isolation is around 55 dB.
The power splitter allows us to simultaneously send in a weak tone near the JPA resonance as a way to monitor and stabilize the JPA gain \cite{HAYSTAC2017}. The variable temperature stage (VTS) allows us to carry out {\it in situ} noise calibration \cite{Brubaker2017procedure} as part of the power calibration for the axion signal injected. The Haloscope At Yale Sensitive To Axion Cold dark matter (HAYSTAC) has been using squeezed-state receivers since 2021~\cite{HAYSTAC2021, haystac2023}: the squeezing JPA (SQZ) is also included in this diagram, as it affects the total noise, even though the injected signal does not go through it; the full scheme with additional details for the squeezed-state receiver chain can be found in Refs.~\cite{HAYSTAC2021, haystac2023, backes2021}.

To synthesize the microwave signal for injection, we use the ``Step (digital) sweep'' mode of PSG2 by providing it with a list of frequencies, amplitudes, and dwell times. In aggregate, the list of rf tones would reproduce the axion's lineshape. In this experiment, by default, we fix the amplitude and dwell time and only change the frequency of each point. The frequency precision is set to 0.1 kHz or 8 significant digits for a GHz frequency, chosen based on the precision of data acquisition. Due to the device's output-byte limit during transmission, the maximum length for such a list is about $n_8\!=\!35$. For sub-kHz frequency precision, we empirically found that $N\!\gg$2,000 samples are needed to reveal the shape of the sampling distribution. Therefore, to iterate through a list with $N \!\gg \!n_8$ rf tones, it is necessary to transmit $\lceil N/n_8 \rceil$ sublists. To begin, we establish the connection between PSG2 and PC2 via virtual instrument software architecture (VISA) and  enable the rf output on PSG2. 
The steps for sending $N$ rf tones are as follows:
\begin{enumerate}
    \item Send $n_8$ frequencies, dwell time $\tau_d$, and amplitude $P_\text{syn}$ to PSG2 from PC2 in one command.
    \item Wait for $\tau_w$ before sending another command to PSG2. Here $\tau_w= n_8\tau_d+ \tau_r$ includes the time required for the sweep $n_8\tau_d$ and an additional uniformly random delay between 2 and 5 seconds $\tau_r\sim U(2,5)$ required for VISA communication. 
    \item Check PSG2's status to confirm that the sweep completed. If not, generate an error message reporting PSG2's status (settling or sweeping) and wait 2 more seconds before checking the status again.
    \item Repeat steps 1-3 until all $N$ items are iterated over.
\end{enumerate}
Based on the timing parameters, we can derive the average duty cycle of this signal injection routine as $\bar \eta=n_8\tau_d/\bar \tau_w$, after averaging over $\lceil N/n_8 \rceil\gg$1 repetitions. By default, the dwell time is $\tau_d=\! 5$ ms, resulting in $\eta =\!4.8 \% \pm \frac{1 \%}{\lceil N/n_8 \rceil}$. 
In step 2, $\tau_r$ is randomized to avoid introducing any additional frequency patterns into the signal.

\subsection{Rejection sampling \label{subsec: rejectionsampling}}
\begin{figure}[h]\centering
\includegraphics[width=\columnwidth]{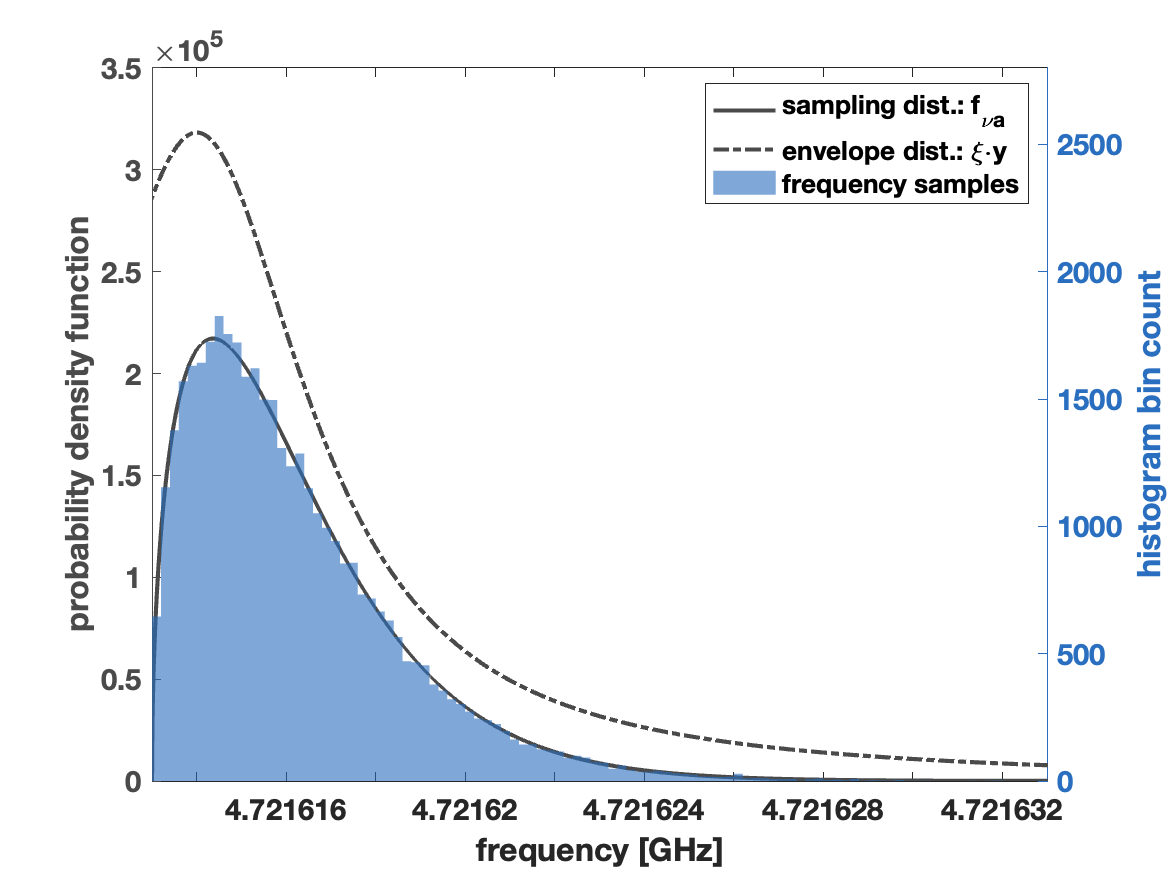} 
\caption{Applying rejection sampling to create a list of $N\!=\!40,000$ random frequencies from an axion's spectral distribution with $\nu_a$=4.7216300 GHz. The envelope distribution (dash-dotted gray line) for the sampling distribution $f_{\nu_a}$ (solid gray line) is obtained by scaling the Cauchy distribution $y$ by $\xi=$3. The histogram (blue) shows the samples obtained using this method conform to the sampling distribution $f_{\nu_a}$. 
  }\label{fig: rejectionsampling}
\end{figure}

To generate a list of random frequencies for signal injection, the probability density function of the sampling distribution is the axion's spectral function
\begin{widetext}
    \begin{align}
        f_{\nu_a}(\nu) &= \frac2{\sqrt\pi} \left(\sqrt{\frac32} \frac1r \frac1{\nu_a\braket{\beta^2}}\right) \sinh\left( 3r\sqrt{\frac{2(\nu-\nu_a)}{\nu_a\braket{\beta^2}}}\right) e^{-\frac{3(\nu-\nu_a)}{\nu_a\braket{\beta^2}}-3r^2/2},\label{eq: f}
    \end{align}
\end{widetext}
where $\sqrt{\braket{v^2}} \approx$270 km/s is the virial velocity, $\braket{\beta^2} =\braket{v^2}/c^2\approx\! 8\times 10^{-7}$ ($c$ is the speed of light), $v_s\approx\! 220$ km/s the orbital velocity of the solar system about the center of the galaxy, and $r=v_s/\sqrt{\braket{v^2}}\approx\!\sqrt{2/3}$ \cite{Brubaker2017procedure}. Given the function form of $f_{\nu_a}$, it can be seen that applying inverse transform sampling to sample from it is not obviously trivial. Here we use rejection sampling \cite{Flury} as an alternative. To apply it, we find another distribution whose probability density function $y(\nu)$ satisfies $\xi \cdot  y(\nu)>f_{\nu_a}(\nu) \: \forall \nu$ for some $\xi >$0. In our case, the Cauchy distribution
\begin{align}
   y(\nu) = \frac{1}{\sigma \pi} \frac{1}{1+((\nu-\mu)/\sigma)^2}
\end{align}
with $\mu =\!\nu_a +1$ kHz and $\sigma$=3 kHz, is sufficient because as $\nu\rightarrow \infty$, $f_{\nu_a}/y\rightarrow 0$. %\footnote{In MATLAB, the software we use, a Cauchy distribution can be obtained by modifying the t-location scale distribution---fixing the shape parameter to 1}.
Furthermore, the cumulative function of a Cauchy distribution is invertible; therefore it is straightforward to apply inverse transform sampling to sample from $y$. 
Let us consider rejection sampling as an iterative process. In each iteration, we sample a random number from $y$, noting it as $\nu'$, and then decide whether to reject or accept it based on another random number, $u$, chosen from the uniform random distribution between 0 and 1. If $u \le f_{\nu_a}(\nu')/(\xi\cdot y(\nu'))$, we accept $\nu'$; otherwise reject. This iteration continues until we have $N$ samples. 
Fig.~\ref{fig: rejectionsampling} shows the outcome of this method when the sampling distribution is $f_{\nu_a}$ with $\nu_a$=4.7216300 GHz as an example. It can be seen that the overall acceptance rate depends on the ratio of $\int d\nu f_{\nu_a}$ to $\int d\nu \: \xi\cdot y$, i.e. the ratio of areas under $f_{\nu_a}$ and $\xi\cdot y$; with $\xi=$3, $\int d\nu f_{\nu_a}/\int d\nu \: \xi\cdot y\approx 0.3$. 

\subsection{Validation of Spectral Shape}
\begin{figure}[h]   
   \centering
   \includegraphics[width=\columnwidth]{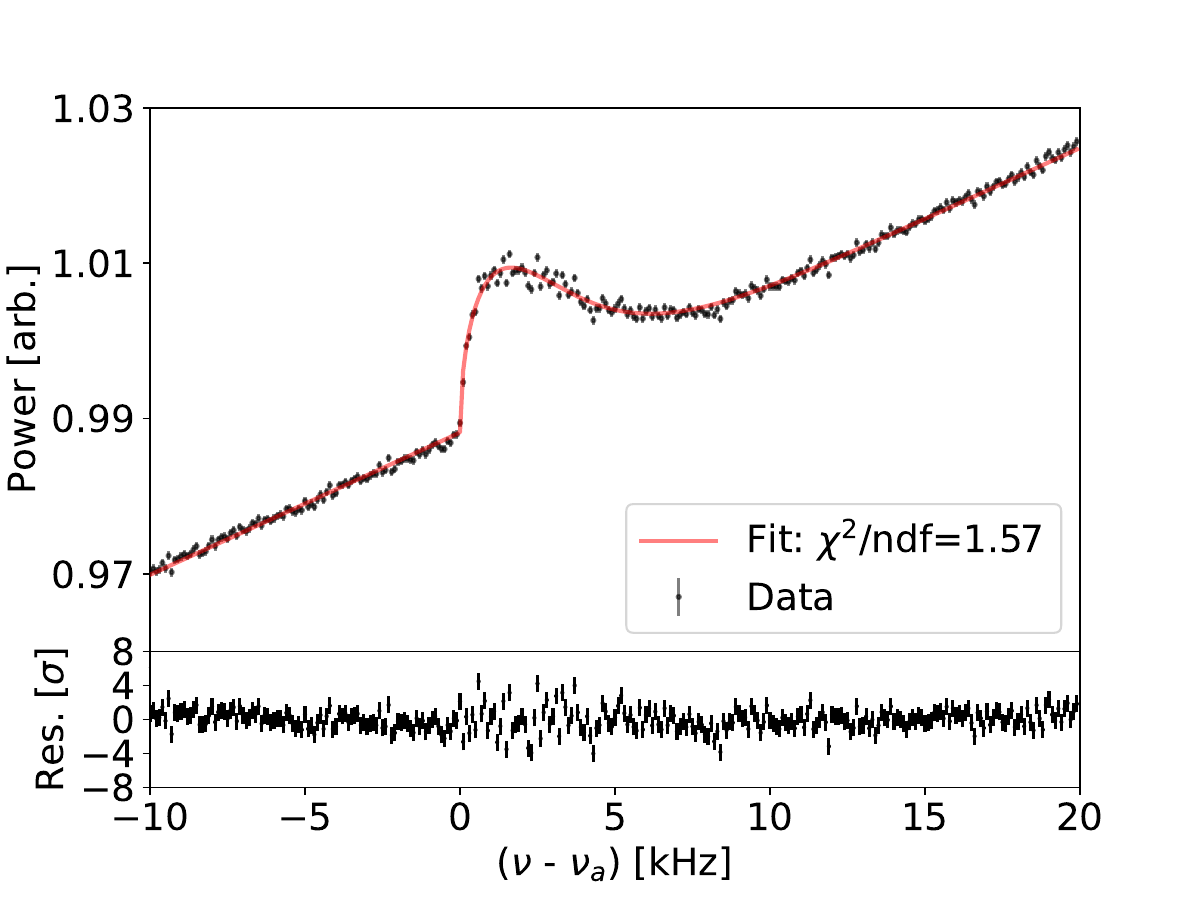}
   \caption{
   % \textcolor{red}{
   (top) Observed power in a 30~kHz window near an injected signal from a high power injection run.  To validate the signal shape, the spectrum is minimally processed with no filtering applied.  A fit to the lineshape in Eq.~\ref{eq: f} is shown as a solid line.  This fit includes a linear~($m\nu + b$) background component to capture the cavity's spectral variation over this range. The residuals between the fit and data are shown in the bottom panel and give a $\chi^{2}/\mathrm{ndf}$=1.57 showing good agreement between the observed and injected spectral shapes.
   % }
   }
   \label{fig:shape_validation}
\end{figure}

% \textcolor{red}{
Validation of this procedure is performed with a $\sim$10 hour signal injection run to verify the signal has the correct spectral shape.  The injection is done at relatively high power such that the signal is clearly visible above the noise even without the standard processing~\cite{Brubaker2017procedure}.  This decouples any deviations caused by the injection routine from shape changes introduced by the  processing and filtering scheme used to analyze the data. The observed signal is fit to the expected shape given by Eq.~\ref{eq: f} in a 30~kHz window around the injected frequency.  The fit model includes a linear background component to approximately capture the spectral shape of the cavity in this range.  Results from the fit in which only the amplitude of the expected signal per Eq.~\ref{eq: f} and the background model parameters are floating are shown in Fig.~\ref{fig:shape_validation}.  This results in a $\chi^{2}/\mathrm{ndf}$ of 1.57 showing good agreement between the observed and injected signal shapes.  The $\chi^{2}/\mathrm{ndf}$ can be improved minimally to 1.56 when allowing the width of the lineshape to float by also varying the virial velocity but remains within 1$\%$ of the true velocity used in the injection.
% }

\subsection{Power calibration 
\label{subsec: calibration}}
\begin{figure}[h]    
    \centering
    \includegraphics[width=\columnwidth]{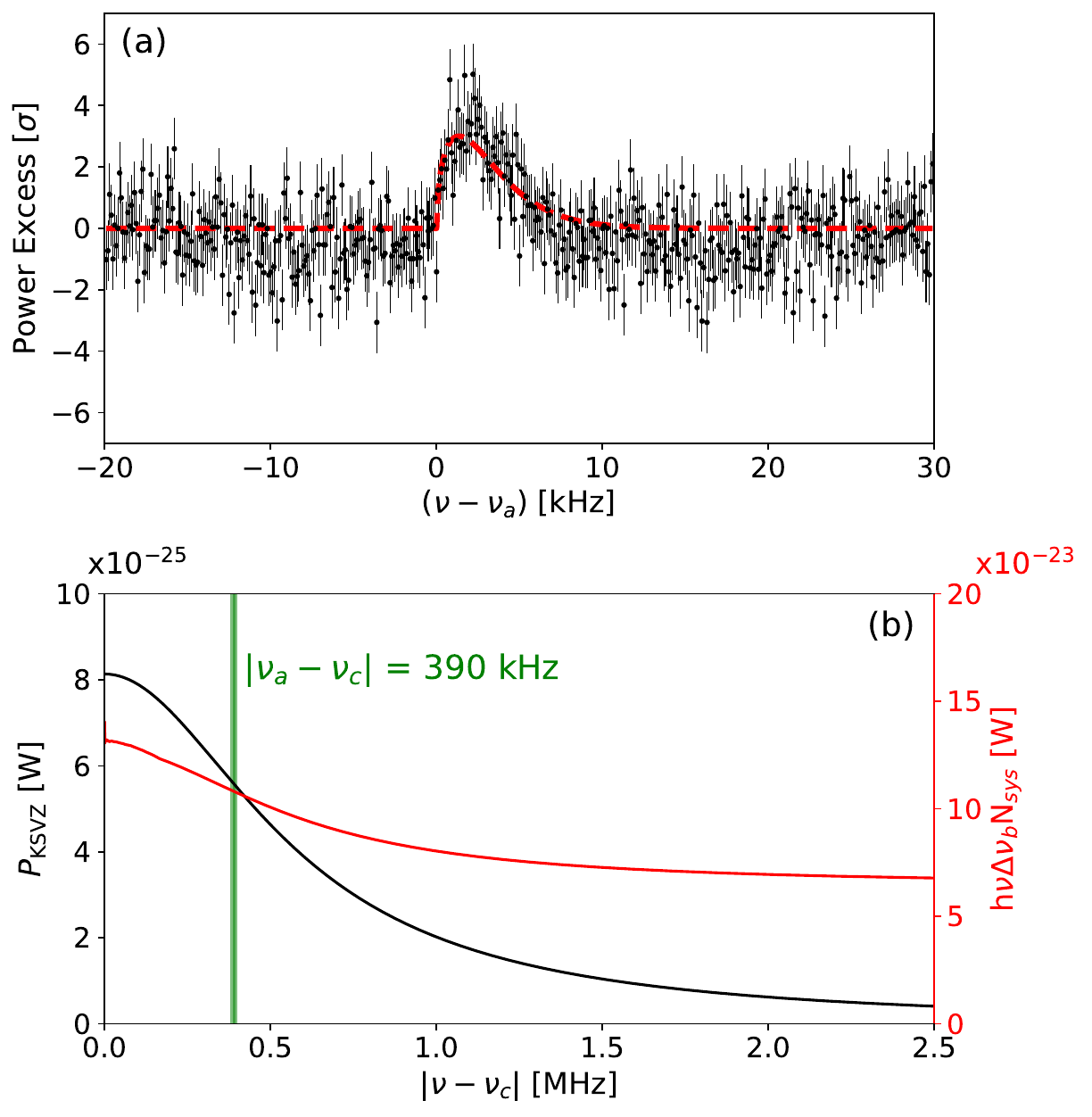}
    \caption{(a) Power excess (points) in the 
    % \textcolor{red}{
    combined spectrum normalized to the expected standard deviation (error bars)
    % }
    and overlaid with $f_{\nu_a}$ (dashed line) in a 
    % \textcolor{red}{
    50~kHz
    % }
    window around the synthetic axion signal . (b) Expected signal power $P_\text{KSVZ}$ (black) and system noise $h\nu_c N_\text{sys}$ integrated over an analysis binwidth $\Delta \nu_b$ (red) at fixed cavity frequency $\nu_c$ as functions of detuning $|\nu-\nu_c|$. 
    % \textcolor{red}{
    The green shaded region indicates the frequencies within $\Delta \nu_{a}$ of the injected signal.
    % }
    The cavity presents a noise source to the receiver that is frequency dependent with $\sim1.1$ MHz bandwidth and varies with $|\nu-\nu_c|$. The response to an axion signal has a different frequency dependence that is more sharply peaked at the cavity resonance. The plot shows the detuning-dependent variation of the axion signal and system noise. }
    \label{fig:calib}
\end{figure}

%\begin{figure}[h]    \centering
% \begin{subfigure}{.5\textwidth}\centering 
% \caption{}\label{fig: powerexc}
%    \includegraphics[width=\columnwidth]{powerexcessfakeaxion.eps}
%    \caption{}\label{fig: factor}
%    \includegraphics[width=\columnwidth]{pksvzNsyslatestW.eps}
%\end{subfigure}
%    \caption{(a) Power excess in a rescaled spectrum overlaid with $f_{\nu_a}$ and zoomed in on the synthetic axion signal (dashed line). (b) Expected signal power $P_\text{KSVZ}$ (gray) and system noise $h\nu_c N_\text{sys}$ integrated over an analysis binwidth $\Delta \nu_b$ (orange) at fixed cavity frequency $\nu_c$ as functions of detuning $|\nu-\nu_c|$. The green dashed line indicates the frequency of the injected signal. The cavity presents a noise source to the receiver that is frequency dependent with $\sim1.1$ MHz bandwidth and varies with $|\nu-\nu_c|$. The response to an axion signal has a different frequency dependence that is more sharply peaked at the cavity resonance. The plot shows the detuning-dependent variation of the axion signal and system noise. }
%    \label{fig: calib}
%\end{figure}

Though the power of each rf tone injected can be programmed on PSG2, neither the weak port's coupling efficiency nor the transmission line loss is known, so it is necessary to carry out a power calibration measurement to determine the time-averaged power of a synthetic axion signal. At fixed cavity frequency, suppose we inject axions, each with power $P_\text{syn}$, near the cavity resonance for duration $\tau_\text{int}$. The observed signal-to-noise ratio (SNR) depends on the system noise power at the injected axion's detuning with respect to the cavity frequency $\nu_c$, $h\nu_c N_\text{sys}(|\nu_a-\nu_c|)$ ($h$ is the Planck constant), $P_\text{syn}$, $\tau_\text{int}$, $\eta$, and the measurement bandwidth which is chosen to match the axion linewidth $\Delta \nu_a\sim \nu_a/10^6$. This then allows us to estimate $P_\text{syn}$ from a standard axion measurement  
\begin{equation}
    P_\text{syn}=\text{SNR} \cdot (h\nu_c N_\text{sys}(|\nu_a-\nu_c|)) \cdot\sqrt{\frac{\Delta \nu_a}{\eta \tau_\text{int}}}.
\label{eq: ii3}
\end{equation} 
As an example, Fig.~\ref{fig:calib} shows some results from a 12-hour measurement with the cavity and JPA at fixed frequencies. 
% \textcolor{red}{
This data is taken at a lower injected power to produce a more realistic signal which is closer to the detection threshold of the experiment
% }. 
From this dataset, SNR and $N_\text{sys}$ are derived following the analysis routine outlined in Ref.~\cite{Brubaker2017procedure}. 
Each raw spectrum is first normalized by dividing out the average baseline. This step allows us to identify bins contaminated by IF noise. After removing contaminated bins from the first normalized spectrum, the spectrum is normalized again by dividing out the Savitzky-Golay fit to itself and subtracting 1. Then the resulting spectrum are rescaled to account for the detuning-dependent sensitivity 
% \textcolor{red}{
and summed to produce a combined spectrum.  This spectrum is normalized to the expected standard deviation in each bin as shown in Fig.~\ref{fig:calib}a.
% }
To derive the SNR of the injected signal, we find the total power of the injected signal by convolving the spectrum in Fig.~\ref{fig:calib}a with the axion lineshape function $f_{\nu_a}$. In this case, the resulting SNR is about 17.6. The scale factor used in producing the rescaled spectrum is the ratio between the noise power per analysis bin $\Delta \nu_b$=100 Hz and expected signal power, both of which are plotted in Fig.~\ref{fig:calib}b. At detuning $|\nu_a-\nu_c| \sim$390~kHz, the power of an axion-converted microwave signal is expected to be $P_\text{KSVZ} =\! 5.6(1)\times 10^{-25}$~W, assuming the axion-photon coupling strength $g_{a\gamma\gamma}$ (appearing in the axion-photon Lagrangian) in the Kim-Shifman-Vainshtein-Zakharov (KSVZ) model~\cite{DINE1981, Kim1979}, $g_{a\gamma\gamma}^\text{KSVZ}=\!(-3.70\times 10^{-7} \text{/MeV}^2)\cdot m_a$ nominally \cite{HAYSTAC2017}.  
This is calculated from various experimental parameters, including the magnetic field strength and cavity volume~\cite{Sikivie1985, HAYSTAC2021}. The system noise in units of photon numbers $N_\text{sys}(|\nu_a-\nu_c|)$=0.26(2) is derived from {\it in situ} calibration measurements~\cite{HAYSTAC2021, haystac2023, backes2021}. 
Evaluating Eq.~\ref{eq: ii3} with these values, we find $P_\text{syn}\!=\!2.2(2)\times 10^{-23}$~W, and that the injected axions have a coupling strength at $6.2(5) \: g_{a\gamma\gamma}^\text{KSVZ}$ level. 

\section{Summary \label{sec: outlook}}
We demonstrated synthetic axion injection using the rf hopping method \footnote{The code for implementing synthetic axion injection is made available as an open-source project at \href{https://github.com/yuqizhuyqz/syntheticaxioninjection.git}{https://github.com/yuqizhuyqz/syntheticaxioninjection.git}}. The injected axion has the spectral distribution of a cold dark matter axion and its power is calibrated using the total noise power as a reference. This is a more realistic synthetic signal, in terms of spectral shape and power, as compared to the prior demonstration in HAYSTAC Phase I (cf. Appendix F in Ref.~\cite{Brubaker2017procedure}). Furthermore, as HAYSTAC and other axion haloscope experiments, including CAPP and ADMX, enter the data production phase, mitigating bias will become increasingly important \cite{Baxter2021}. For instance, salting and blinding have become an integral part of WIMP (weakly interacting massive particle) dark matter searches \cite{lux2017, XENON1T}. The method we developed can serve as the basis for implementing blinding and salting for axion dark matter searches.

\begin{acknowledgments}
We acknowledge the full HAYSTAC collaboration for the apparatus we used in this experiment, including the microwave cavity and JPAs. Y.Z. thanks Joseph Howlett for helpful discussion and references on mitigating bias in dark matter searches. HAYSTAC is supported by the National Science Foundation under grant numbers PHY-1701396, PHY-1607223, PHY-1734006 and PHY-1914199 and the Heising-Simons Foundation under grants 2014-0904 and 2016-044. Y. Z., S. G., E. G., S. C., M. J., and R. M. are supported in part by the Department of Energy under Grant No. DE-AC02-07CH11359. S. G. is supported in part by the National Science Foundation under Grant No. DMR-1747426. 
\end{acknowledgments}

\section*{AUTHOR DECLARATIONS}
\subsection*{Conflict of Interest}
The authors have no conflicts to disclose.

\subsection*{Author Contributions}
Yuqi Zhu: Investigation (equal); Data Curation (equal); Formal analysis (equal); Writing--original draft (lead).
M. J. Jewell: Investigation (equal);  Data Curation (equal); Formal analysis (equal); Writing---Review \& Editing (equal). 
Claire Laffan: Data Curation (equal); Formal analysis (equal). 
Xiran Bai: Writing---Review \& Editing (equal). 
Sumita Ghosh: Writing---Review \& Editing (equal). 
Eleanor Graham: Writing---Review \& Editing (equal).
S. B. Cahn: Writing---Review \& Editing (equal).
Reina H. Maruyama: Project Administration (equal); Funding acquisition (equal); Writing---Review \& Editing (equal). 
S. K. Lamoreaux: Project Administration (equal); Funding acquisition (equal); Conceptualization(lead); Writing---Review \& Editing (equal). 

\section*{Data Availability Statement}
The data that support the findings of this study are available from the corresponding author upon reasonable request.

\bibliography{references}

\end{document}